\documentclass[10pt,letterpaper]{article}
\usepackage{opex3}
\usepackage{graphicx}

\begin{document}

%
%

\title{Ultra-low power generation of twin photons in a compact silicon ring resonator}
\vspace{-10pt}
\author{Stefano Azzini$^1$, Davide Grassani$^1$, Michael J. Strain$^2$, Marc Sorel$^2$, L.G. Helt$^3$, J.E. Sipe$^3$, Marco Liscidini$^1$, Matteo Galli$^1$, Daniele Bajoni$^{4,*}$ }

\address{
$^1$Dipartimento di Fisica, Universit\`{a} degli Studi di Pavia, via Bassi 6, Pavia, ITALY
\\
$^2$School of Engineering, University of Glasgow, Glasgow G12 8LT, UK \\
$^3$Department of Physics and Institute for Optical Sciences, University of Toronto, 60 St. George St., Toronto, ON M5S1A7, CANADA \\
$^4$Dipartimento di Ingegneria Industriale e dell'Informazione, Universit\`{a} degli Studi di Pavia, via Ferrata 1, Pavia, ITALY}

\email{daniele.bajoni@unipv.it} 



\begin{abstract} 
We demonstrate efficient generation of correlated photon pairs by spontaneous four wave mixing in a 5 $\mu$m radius silicon ring resonator in the telecom band around 1550 nm. By optically pumping our device with a 200 $\mu$W continuous wave laser, we obtain a pair generation rate of 0.2 MHz and demonstrate photon time correlations with a coincidence-to-accidental ratio as high as 250. The results are in good agreement with theoretical predictions and show the potential of silicon micro-ring resonators as room temperature sources for integrated quantum optics applications. 
\end{abstract}



\section{Introduction}
The study of quantum states of light is a rapidly developing field of research. Already exploited for fundamental research in quantum physics, the ability to produce states showing correlations that cannot be explained in a classical way is becoming a fundamental resource for applications. These include quantum cryptographic protocols for communications, secured using quantum uncertainty rather than classical protocols \cite{Gisinreview}, and quantum computing \cite{Eckertreview}.

Nonclassical states of light are usually produced using optical nonlinearities in bulk crystals, in particular parametric fluorescence \cite{boyd_book,Aspect,Kwiat}. Major research efforts are focused on integrating a source of quantum states of light with semiconductor photonic circuits \cite{Berger,Weihs}, particularly on silicon based platforms\cite{TakesueWG,TakesueEPR}.
If realized on a silicon on insulator (SOI) structure, such a source could be integrated with a number of recently developed components for routing, switching and processing of photonic signals \cite{Almeida1,Tanabe1}, following the paradigm of integrated electronic circuits.

Silicon ring resonators \cite{Almeida1,lipson08} are excellent candidates for such sources. In fact, they can have several modes in the telecom band with extremely sharp line widths (on the order of tens of $\mu$eV) and they are fabricated using top down approaches \cite{CMOS1,CMOS2}  compatible with CMOS technology \cite{CMOS3,CMOS4}. Furthermore, they are well suited to optical fiber butt-coupling. These structures are an industrial standard already used in photonics as filters due to their narrow resonances and have already been shown to have strong optical nonlinear properties\cite{Almeida1,tuner08}.

In this paper we show efficient emission of correlated photon pairs from a silicon ring resonator with a 5 $\mu$m radius, i.e. a footprint of only 80 $\mu m^2$. The photon pairs are generated in two separate spectral bands by spontaneous four wave mixing \cite{Helt} using a continuous wave laser pump, and are detected using single photon counting modules. We show a strong coincidence peak in time-correlated measurements, with a coincidence to accidental ratio comparable to the best results obtained in structures with a much larger footprint \cite{TakesueWG}, and an order of magnitude larger than the only preceding report on a silicon ring resonator \cite{clemmen09}.

This paper is organized as follows: in section 2 we present details about the sample fabrication and characterization, along with results on spontaneous four wave mixing. In section 3 we report on coincidence measurements between the generated signal and idler beams and discuss the feasibility of this device as an efficient source of nonclassical photonic states. Finally, in section 4 we summarize our results and conclusions.

\section{Sample characterization and spontaneous four wave mixing}

\begin{figure*}
\centerline{\includegraphics[width=0.95\columnwidth]{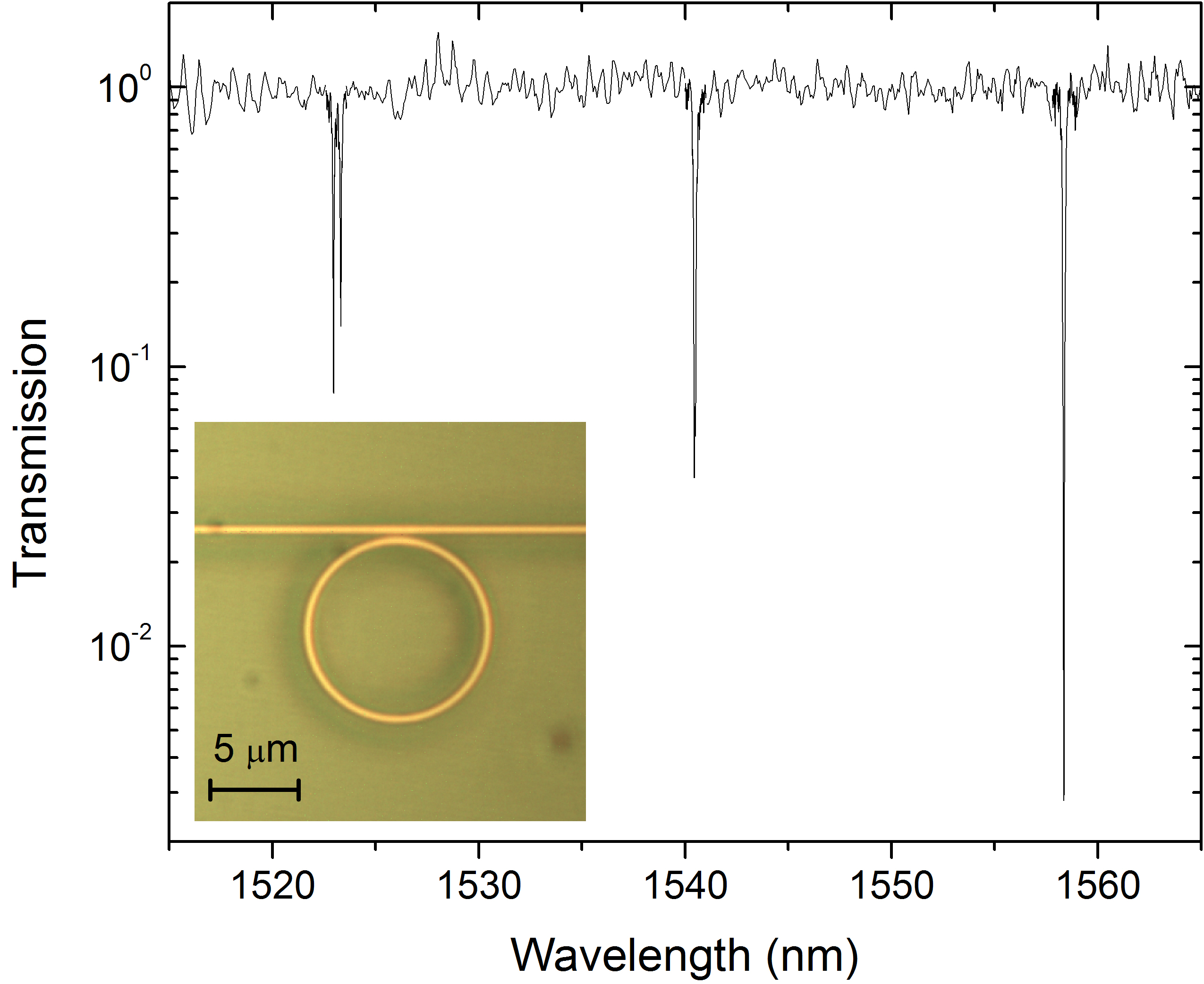}}
\caption{Transmission spectrum from the sample. The inset shows an optical microscope image of the ring resonator.}
\label{Fig1}
\end{figure*}

The device employed in our experiments is a compact silicon micro-ring resonator. It was fabricated using state-of-the-art electron beam lithography techniques followed by inductively coupled plasma reactive ion etching on a silicon-on-insulator planar waveguide with a 220 nm silicon core on a 2 $\mu$m thick silica cladding. The ring is circular with a radius of 5$\mu$m (an optical microscope image is shown in the inset of Fig. \ref{Fig1}) and is evanescently coupled to a silicon nano-wire waveguide via a deep-etched 125 nm gap point coupler. This is to good approximation the critical-coupling distance, designed in order to transfer the maximum amount of light into the micro-resonator, as well as provide the best filtering action of the incoming light beam by the device. Transverse dimensions of both the ring and the waveguide are 500 nm x 220 nm. Efficient input and output coupling to the narrow waveguide is of fundamental importance for performing low-pump-power nonlinear optics experiments, and it was achieved using spot-size converters \cite{SSC1,SSC2}: 300 $\mu$m long silicon inverse tapers ending in a 20 nm tip width covered by 1.5x2.0 $\mu$m$^{2}$ polymer waveguides. The cavity was designed in order to offer a compromise between a high quality factor, a free spectral range (FSR) allowing for feasible filtering, and a small foot-print.

Linear transmission spectra with a resolution of 5 pm were taken using a tunable continuous wave laser modulated through an optical chopper for injection and an InGaAs photodiode coupled to a lock-in amplifier for detection. The three modes we are considering are shown in Fig.  \ref{Fig1}: the free spectral range is about 20 nm and the quality factor is $Q\simeq$7900. The wavelengths of the three resonances are $\lambda_i=1558.35$ nm, $\lambda_p=1540.5$ nm, while the resonance around 1523 nm is split at 1523 nm and 1523.3 nm. Double resonances are a well known feature of micro-ring resonators, and they originate from the coupling of counter-propagating modes induced by surface roughness \cite{Little}.

The nonlinear process used to generate photon pairs in our sample is spontaneous four wave mixing. In this process, two photons injected by the laser pump at energy $E_p$ (corresponding in our sample to the resonance at 1540.5 nm) are destroyed, and two photons are created, one at energy $E_s$ (signal) and one at energy $E_i$ (idler). To be efficient this process must conserve both the total energy and momentum and it is amplified whenever pump and generated photon energies are simultaneously tuned to the resonances of the ring.

\begin{figure*}
\centerline{\includegraphics[width=\columnwidth]{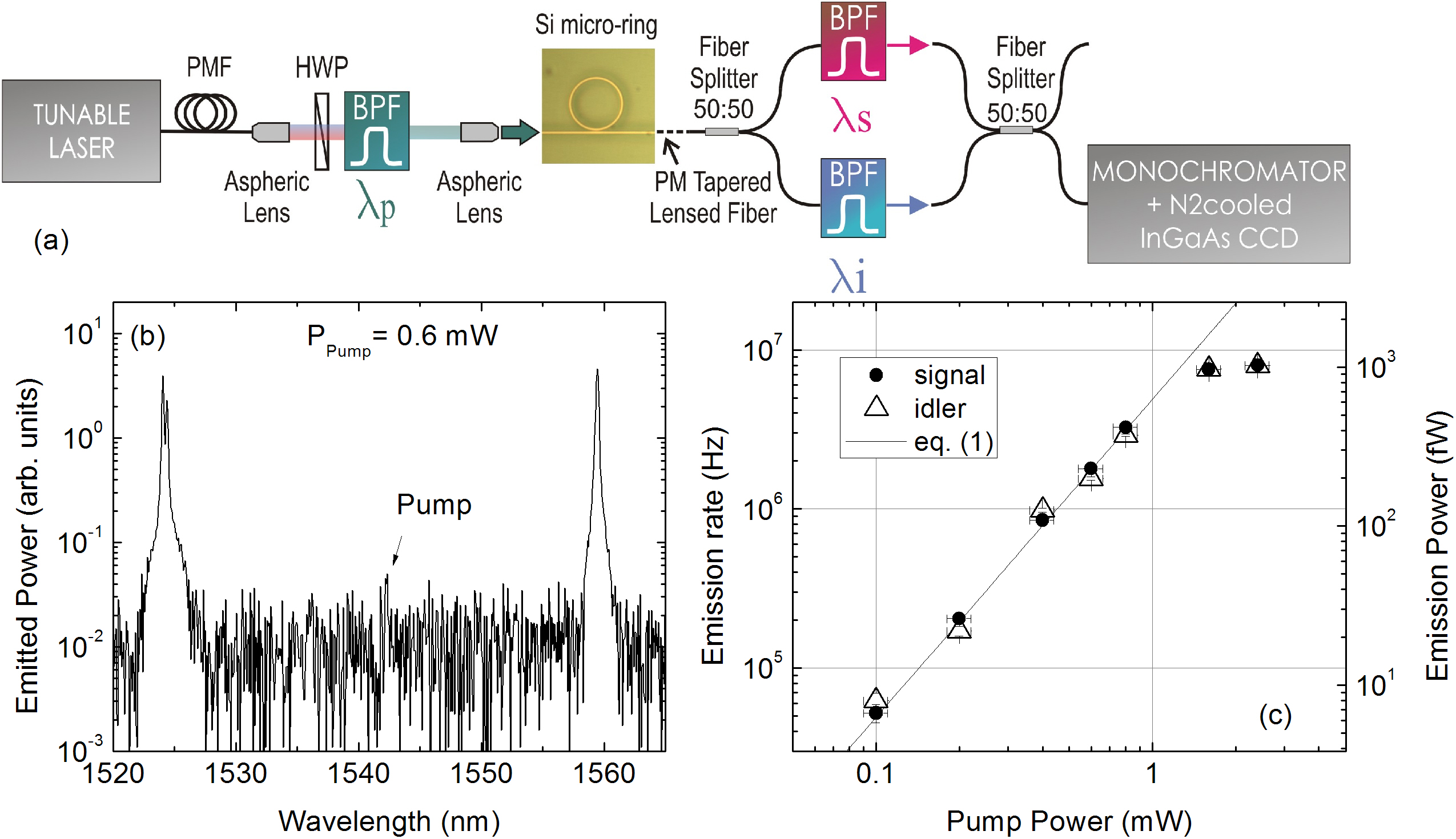}}
\caption{(a) Schematic of the experimental setup used to test spontaneous four wave mixing. (b) Spectrum of spontaneously generated signal and idler beams for $P_{Pump}=0.6$ mW. (c) Integrated intensities of the generated beams as a function of the pump power. The line is the prediction from eq. (1).}
\label{Fig2}
\end{figure*}

The experimental setup used to test spontaneous four wave mixing (sFWM) is sketched in Fig.\ref{Fig2} (a). Light from a CW laser (Santec,TSL-510) tunable across the C-band (1500-1630 nm) comes through a polarization-maintaining (PM) single-mode fiber cable at the pump wavelength of $\sim$1540.5 nm. Spectral filtering of the laser beam is performed before injection to filter out the fluorescence background emitted from the laser because of amplified spontaneous emission: our set-up achieves more than 120 dB of side-band suppression in the spectral regions where signal and idler photons are going to be generated by the sFWM process. The pump beam is collimated to free-space using an aspheric lens, its polarization is aligned to the TE-like single-mode of the silicon waveguide, and finally the pump light is coupled to the sample using an aspheric lens. The output from the sample is collected with a PM lensed-fiber mounted on a high-precision fiber rotator. The total insertion losses of the silicon chip are measured to be about 7 dB, which are equally distributed before and after the ring resonator. The output containing the transmitted pump beam and the photon pairs is sent, through a 50:50 fiber beam-splitter, to two band pass filters (BPF) tuned to the signal and idler frequencies in order to reject the remaining pump. We estimate a total loss of 13 dB from the ring to the output of the filters. Using a 50:50 fiber beam-combiner, the output of both BPFs is finally conveyed to a monochromator coupled to a nitrogen-cooled CCD. 

An example of a sFWM spectrum is shown in Fig. 2 (b). Two emission peaks are found in correspondence with the signal and idler resonances. The residual of the pump beam after the filters is barely visible over the background noise in Fig. 2 (b), and we have verified that the pump residual is always at least 25 dB weaker than the signal and idler beams. This is of fundamental importance for photon counting experiments described in next section, as a strong pump beam would generate a large number of accidental coincidences.

Estimated emission rates for the signal and idler resonances are shown in Fig. 2 (c) as a function of the pump power $P_{Pump}$. The value of $P_{Pump}$ reported in the figure is the internal pump power, extracted from the power at the laser output by considering the losses from the filter and the coupling losses to the sample. The emission rates are estimated  by calibrating the response of the CCD camera using a power meter with pW sensitivity and then integrating the measured emission lines for the signal and idler resonances (notice that the signal resonance is split, and the line is integrated over both peaks). The measured generation rate, as determined by taking into account the system losses is, for example, $R\simeq$0.2 MHz for 0.2 mW of coupled pump power. This generation rate is in excellent agreement with the theoretical curve given by \cite{Helt,note_critical}:
\begin{equation}\label{spontaneous}
\rho=(\gamma 2\pi R)^2\left(\frac{Q v_g}{\omega_p\pi R}\right)^3\frac{v_g}{4\pi R}P_{Pump}^2,
\end{equation}
where, in our case, $v_g=$1.2$\cdot 10^8$ m/s is the group velocity and the nonlinear parameter $\gamma=$190 W$^{-1}$m$^{-1}$ has been estimated by stimulated four wave mixing experiments performed on the same sample \cite{helt12,azzini12}: one roughly gets $\rho$=$5\left(\frac{\textrm{MHz}}{\textrm{mW}^2}\right)P_{Pump}^2$ with $P_{Pump}$ expressed in mW, corresponding to the straight line in Fig. 2 (c). Here the factor $\left(\frac{Q v_g}{\omega_p\pi R}\right)^3$ (about 10$^5$ in our structure) accounts for the overall field enhancement due to the ring resonances.
 
The intensities of the signal and idler beams as reported in Fig. 2 (b) and (c)  are almost equal, and they increase quadratically with $P_{Pump}$: this quadratic dependence is a clear signature of spontaneous four wave mixing. The curve departs from the quadratic dependence and tends to saturate at high pumping power, due to the thermo-optic effect induced by two-photon absorption. Another consequence of this effect is that the resonances slightly redshift with increasing pump power, and the energy of the input laser is retuned accordingly to maximize the output from the sample.

\section{Coincidence measurements}

After having verified the occurrence of sFWM, we proceeded to assess the correlated emission of the photon pairs. The set-up used is almost the same, but now the output from the signal and idler BPFs is sent to two superconducting single photon detectors (SSPDs) as shown in Fig 3 (a). The SSPDs are mounted in a refrigerating unit inside a liquid He dewar at 1.7 K. The SSPDs are biased so that the dark counts are of the order of 100 Hz, and the two detectors have detection efficiencies of 5\% and 10\%.
We use SSPDs because although their detection efficiencies are lower than the typical detection efficiencies of InGaAs avalanche photodiodes, they have a much better time response (65 ps instead of some ns) and much lower rate of dark counts (10$^2$ Hz instead of 10$^4-$10$^5$ Hz). Fiber polarization controllers (FPCs) are used along the two detecting lines in order to maximize the single photon detection rate onto the SSPDs. We estimate the total losses from the ring to the SSPD to be $L_s=24$ dB for the signal beam and $L_i$=27 dB for the idler beam, including the quantum efficiency of the detectors. The temporal resolution of the experiment is limited to 65 ps by the SSPDs. The output from the detectors is correlated using a Picoquant Hydraharp event timer, which records a stream of events on a computer. 
Two examples of coincidence curves elaborated with a coincidence time window corresponding to the response time of the detectors are shown in Fig. 3 (b) and (c) for increasing pumping powers. The curves were taken with an integration time of 20 minutes. The curves show a clear coincidence peak, due to the concurrent emission of signal and idler photons, over a constant background due to accidental events. The width of the peak is given by the response time of the detectors; the linewidth of the signal and idler resonances correspond to a photon coherence time of $\sim$7 ps, much lower the response time of the SSPDs.
The data as a function of the pump power is shown in Fig. 3(d). The total number of coincidences (all the detected coincidences summed over the whole coincidence peak in a $\sim400$ ps time window) increases quadratically with the pump power: as expected, the number of coincidences is proportional to the rate of emission of photon pairs, shown in Fig. 2(c) to be quadratic in $P_{Pump}$. The coincidence rate $\rho_C$ can be directly linked to the pair generation rate $\rho$ through the losses:  $\rho_C=\rho\cdot 10^{-(L_s+L_i)/10}$. By taking the measured rate of $\rho\simeq$0.2 MHz for $P_{Pump}$=0.2 mW, the expected rate of coincidences is $\rho_C\simeq 1.5$ Hz, which is of the same order of magnitude as the measured value of $\simeq$0.5 Hz reported in Fig. 3 (d).

\begin{figure*}
\centerline{\includegraphics[width=\columnwidth]{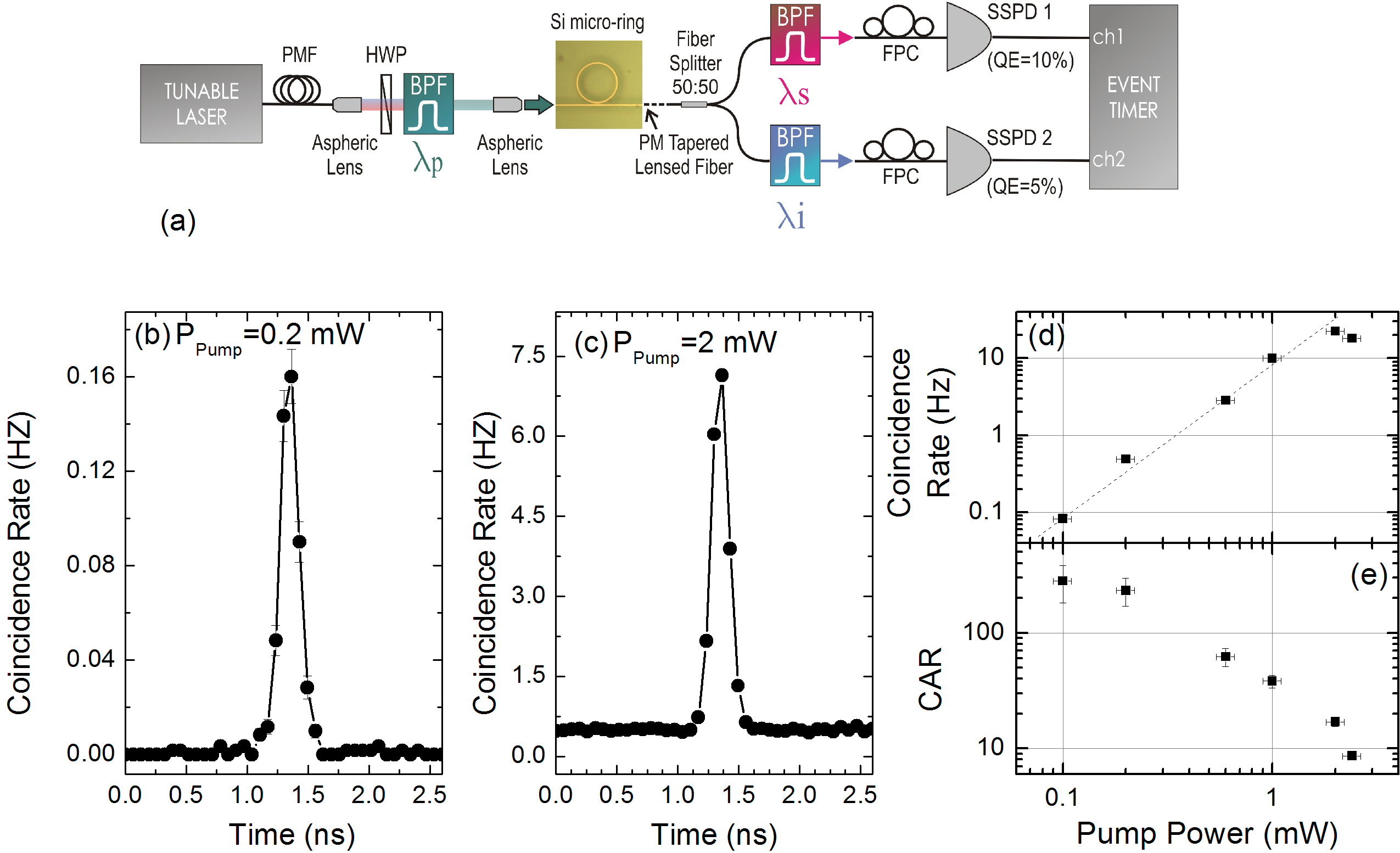}}
\caption{(a) Schematic of the experimental setup used for coincidence measurements. (b) and (c) Coincidence histograms for different pump powers. (d) Rate of coincidences as a function of the pump power. The dashed line is a guide to the eye proportional to the square of the pump power. (e) Coincidences to accidentals ratio as a function of the pump power. }
\label{Fig3new}
\end{figure*}

The coincidence to accidental ratio (CAR) is an important figure of merit to characterize the performance of a source of photon pairs. We calculate the CAR taking the number of coincidences within the time response of the detectors \cite{TakesueWG,clemmen09,davanco12} over the average of the background on the same time window taken apart from the peak. The results as a function of the pump power are shown in Fig. 3 (e). We measure a maximum value of 280 for $P_{Pump}$=0.1 mW. This value, obtained in a device with 5 $\mu$m radius, is comparable to the highest CAR reported in cm-long silicon waveguides \cite{TakesueWG}, it is almost ten times that reported earlier \cite{clemmen09} for a ring resonator, and it is about 25 times that recently demonstrated in a coupled-resonator optical-waveguide \cite{davanco12}. The improved CAR with respect to ref. \cite{clemmen09}, as well as the higher generation rate of signal and higher photons, are due to lower losses in our sample and the better time response of the detectors employed in this work.

Ideally, in the absence of dark counts of the detector, the CAR is expected  to decrease as the inverse of the generation rate \cite{TakesueCAR1}, and thus quadratically with $P_{Pump}$. 
At low powers our generation rate becomes comparable with the dark count rate, meaning that the accidentals are overestimated, thus decreasing the value of the CAR.  As a result, in Fig. 3 (e) the dependence of the CAR on $P_{Pump}$ is no longer quadratic.

Finally, we discuss the feasibility of this device as a source for quantum states of interest, in particular entangled photons or heralded single photons. In the first case, the fact that the pairs are emitted simultaneously can be exploited to achieve time-energy entanglement between signal and idler beams. Experiments to verify entanglement \cite{Franson} rely on double coincidence measurements: the observation of a clear coincidence peak in this work shows that this device could be implemented as a source for such measurement.
Experiments to verify heralded single photons \cite{Whamsley,Razavi} rely on counting triple coincidences; given that the double coincidence rate reported here is of the order of some Hz, triple coincidences may require too long an integration time to pile up a statically significant sample. The problem is mainly due to the losses in the detection line, which could be substantially reduced by direct integration of the optical filters and detectors into the SOI chip \cite{Fiore,nbnsilicon}. Otherwise, using different kinds of resonators, alternative approaches could be used, which do not involve heralding, but generate Fock states via photon-photon repulsion \cite{Gerace}.

\section{Conclusion}
We have studied the efficient emission of correlated photon pairs from a silicon ring resonator, reporting remarkably high values of CAR as well as the emission rate \cite{Thompson}. The observed rates are in excellent agreement with the theoretical prediction. 

This results show that silicon micro-ring resonators are appealing and promising devices for integrated quantum optics applications: they are in fact extremely compact, CMOS compatible and they work at room temperature. Our findings open the route to the demonstration of heralded photon sources and the generation of quantum correlated photonic states in an integrated environment.

\section{Acknowledgments}
We thank L.C. Andreani for careful reading of the manuscript. This work was supported by MIUR funding through the FIRB ``Futuro in Ricerca" project RBFR08XMVY and from the foundation Alma Mater Ticinensis. JES and LGH acknowledge support of the Natural Science and Engineering Research Council of Canada (NSERC).


\begin{thebibliography}{99}
\bibitem{note_critical} Unlike in Ref.\cite{Helt}, here we assume losses in the ring resonator. At the critical coupling, the on-resonance field enhancement in the ring resonator is $FE\simeq\sqrt{2Qv_g/(\omega_0 2 \pi R)}$, with $\omega_0$ the resonant frequency
\end{thebibliography}
\end{document}